\documentclass[twocolumn,aps]{revtex4}

\usepackage{graphicx}     
\usepackage{dcolumn}      
\usepackage{bm}           

\begin{document}

\title{Quantum Error Correction on Linear Nearest Neighbor Qubit Arrays}

\author{Austin G. Fowler$^{1}$, Charles D. Hill$^{2}$ and Lloyd C. L. Hollenberg$^{1}$}

\affiliation{
Centre for Quantum Computer Technology, School of physics\\
$^{1}$University of Melbourne, Victoria 3010, Australia.\\
$^{2}$University of Queensland, Queensland 4072, Australia.\\}
\date{\today}

\begin{abstract}
A minimal depth quantum circuit implementing 5-qubit quantum error
correction in a manner optimized for a linear nearest neighbor
architecture is described. The canonical decomposition is used to
construct fast and simple gates that incorporate the necessary
swap operations. Simulations of the circuit's performance when
subjected to discrete and continuous errors are presented. The
relationship between the error rate of a physical qubit and that
of a logical qubit is investigated with emphasis on determining
the concatenated error correction threshold.
\end{abstract}


\maketitle

The field of quantum computation deals with the manipulation of
2-state quantum systems called qubits. Many different physical
systems are being investigated in the race to build a scalable
quantum computer \cite{Kane98,Enge01,Giov00,Hind01,Jame00,Mooi99}.
Due to the fragility of quantum systems, one property a scalable
architecture must possess is the ability to implement quantum
error correction (QEC) \cite{Shor95,Lafl96,Niwa02}. The question
has been raised as to how well QEC can be implemented on a linear
nearest neighbor (LNN) quantum computer \cite{Gott00} due to the
expectation that numerous swap gates will be required. Working out
a way around this is important due to the large number of LNN
architectures currently under investigation
\cite{Wu99,Vrij00,Gold03,Nova03,Holl03,Tian03,Yang03,Feng03,Pach03,Frie03,Vand02,Soli03,Jeff02,Petr02,Golo02,Ladd02,Vyur00,Kame03}.
In this paper a 5-qubit QEC circuit appropriate for an LNN
architecture is presented that is as efficient as the best known
circuit for an architecture able to interact arbitrary pairs of
qubits \cite{Brau97}.

The paper is organized as follows. Firstly, the canonical
decomposition used to construct efficient 2-qubit gates is
discussed in brief. Details of the method used can be found in
\cite{Krau01}. The Kane architecture \cite{Kane98} has been used
to construct explicit decompositions, but the methods described
apply to any architecture. The 5-qubit QEC scheme is then
discussed and the LNN circuit presented. Following this,
simulations of quantum data storage with and without QEC are
presented. The paper concludes with a summary of all results.

\begin{figure*}
\begin{center}
\resizebox{150mm}{!}{\includegraphics{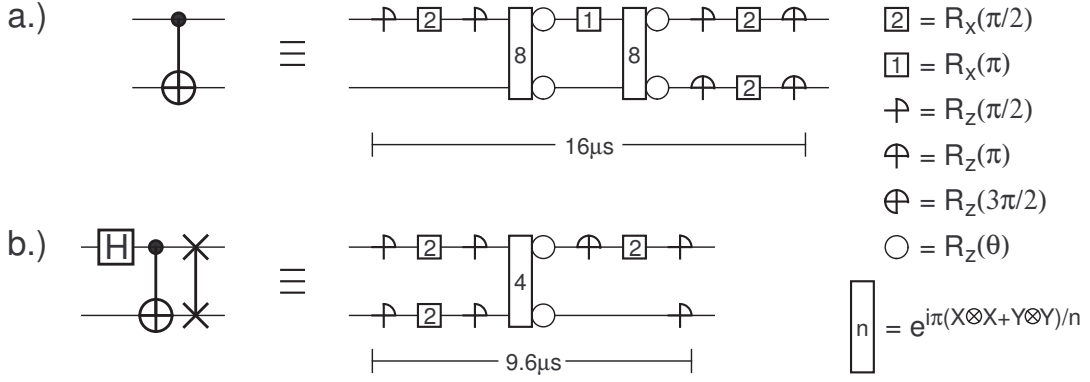}}
\end{center}
\caption{Decomposition into physical operations of a.) CNOT gate
b.) Hadamard, CNOT then swap. Note that the Kane architecture has
been used for illustrative purposes. In addition to the clear
speed advantage when implementing compound gates, the decomposed
CNOT gate is faster than its adiabatic equivalent (26$\mu$s)
\cite{Fowl03}} \label{figure:cnot-hcnots3}
\end{figure*}

The canonical decomposition enables any 2-qubit operator $U_{AB}$
to be expressed (non-uniquely) in the form $V_{A}^{\dag}\otimes
V_{B}^{\dag}U_{d}U_{A}\otimes U_{B}$ where $U_{A}$, $U_{B}$,
$V_{A}$ and $V_{B}$ are single qubit unitaries and
$U_{d}=\exp[i(\alpha_{x}X\otimes X+\alpha_{y}Y\otimes
Y+\alpha_{z}Z\otimes Z)]$ \cite{Krau01}. Moreover, any entangling
interaction can be used to create an arbitrary $U_{d}$ up to
single qubit rotations \cite{Brem02}. These two facts allow the
construction of very efficient composite gates on any physical
architecture. Fig.~\ref{figure:cnot-hcnots3}a shows the form of
such a decomposed controlled-NOT (CNOT) on a Kane quantum computer
\cite{Hill03,Kane98}. The 2-qubit interaction corresponds to
$\alpha_{x} = \alpha_{y} = \pi/8$, and $\alpha_{z} = 0$.
Z-rotations have been represented by quarter, half and
three-quarter circles corresponding to $R_{z}(\pi/2)$,
$R_{z}(\pi)$, and $R_{z}(3\pi/2)$ respectively. Full circles
represent Z-rotations of angle dependent on the physical
construction of the computer. Square gates 1 and 2 correspond to
X-rotations $R_{x}(\pi)$ and $R_{x}(\pi/2)$.
Fig.~\ref{figure:cnot-hcnots3}b shows an implementation of the
composite gate Hadamard followed by CNOT followed by swap
(HCNOTS). Note that the total time of the compound gate is
significantly less than the CNOT on its own.

The implication of the above is that the swaps inevitably required
in an LNN architecture to bring qubits together to be interacted
can be incorporated into other gates without additional cost.
Indeed, in certain cases LNN circuits built out of compound gates
are actually faster. With careful planning, general quantum
circuits can be implemented on an LNN architecture with
asymptotically the same number of gates as that required on an
architecture that allows any pair of qubits to be interacted.

\begin{figure*}
\begin{center}
\resizebox{150mm}{!}{\includegraphics{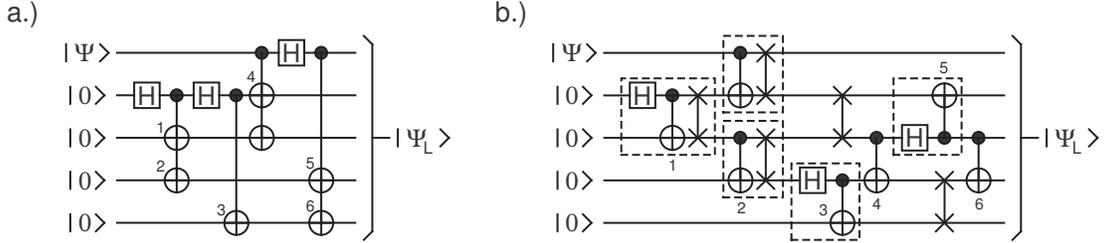}}
\end{center}
\caption{a.) 5-qubit encoding circuit for general architecture b.)
equivalent circuit for linear nearest neighbor architecture with
dashed boxes indicating compound gates. CNOT gates that must be
performed sequentially are numbered.} \label{figure:5qecboth2}
\end{figure*}

\begin{figure*}
\begin{center}
\resizebox{150mm}{!}{\includegraphics{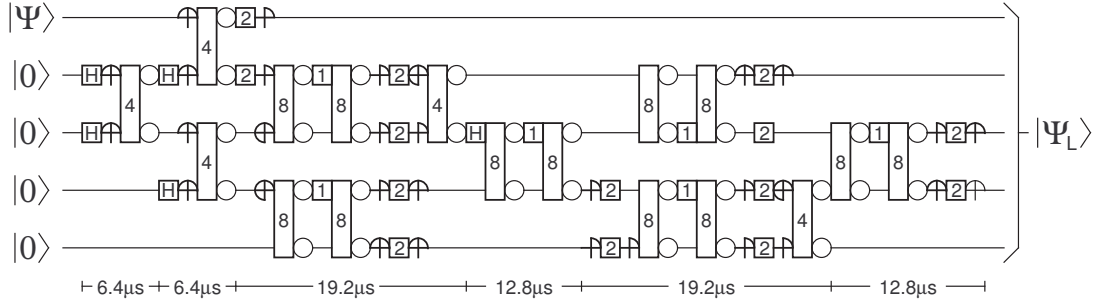}}
\end{center}
\caption{A sequence of physical gates implementing the circuit of
fig.~\ref{figure:5qecboth2}b. Note the Kane architecture has been
used for illustrative purposes.}
\label{figure:qec5dec}
\end{figure*}

\begin{figure*}
\begin{center}
\resizebox{150mm}{!}{\includegraphics{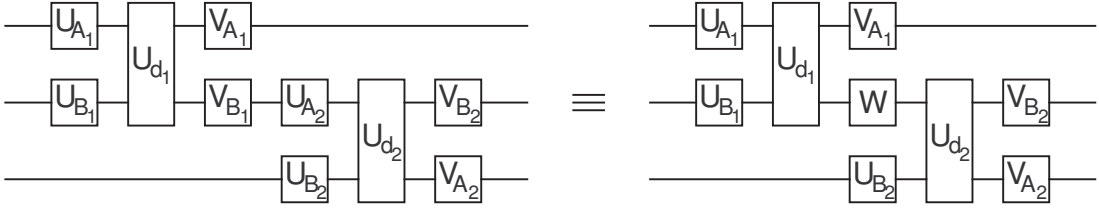}}
\end{center}
\caption{Circuit equivalence used to reduce the number of physical
gates in fig.~\ref{figure:qec5dec}.} \label{figure:equiv}
\end{figure*}

\begin{figure*}
\begin{center}
\resizebox{150mm}{!}{\includegraphics{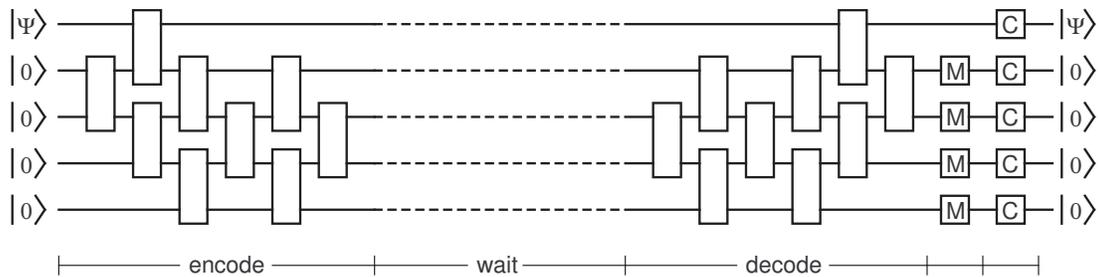}}
\end{center}
\caption{A complete encode-wait-decode-measure-correct QEC cycle.}
\label{figure:cycle}
\end{figure*}

5-qubit quantum error correction schemes are designed to correct a
single arbitrary error. No single error correction scheme can use
less than 5 qubits \cite{Niel00}. A number of 5-qubit QEC
proposals exist \cite{Brau97,Knil01,Lafl96,Niwa02,Benn96}.
Fig.~\ref{figure:5qecboth2}b shows a circuit optimized for an LNN
architecture implementing the encode stage of the QEC scheme
proposed in \cite{Brau97}. For reference, the original circuit is
shown in fig.~\ref{figure:5qecboth2}a. Note that the LNN circuit
uses exactly the same number of CNOTs and achieves minimal depth.
The two extra swaps required do not significantly add to the total
time of the circuit. Fig.~\ref{figure:qec5dec} shows an equivalent
physical circuit for a Kane quantum computer. Note that this
circuit uses the fact that if two 2-qubit gates share a qubit then
two single-qubit unitaries can be combined as shown in
fig.~\ref{figure:equiv}. The decode circuit is simply the encode
circuit run backwards. All 5-qubit QEC schemes are only useful for
data storage \cite{Niwa02} due to the difficulty of interacting
two logical qubits. Fig.~\ref{figure:cycle} shows a full
encode-wait-decode-measure-correct data storage cycle.
Table~\ref{table:one} shows the range of possible measurements and
the action required in each case.

\begin{table}
\begin{tabular}{c|c}
Measurement & Action \\ \cline{1-2}
$\Psi'\otimes$0000 & \texttt{I$\otimes$IIII} \\
$\Psi'\otimes$0001 & \texttt{I$\otimes$IIIX} \\
$\Psi'\otimes$0010 & \texttt{Z$\otimes$IIXI} \\
$\Psi'\otimes$0011 & \texttt{I$\otimes$IIXX} \\
$\Psi'\otimes$0100 & \texttt{I$\otimes$IXII} \\
$\Psi'\otimes$0101 & \texttt{X$\otimes$IXIX} \\
$\Psi'\otimes$0110 & \texttt{Z$\otimes$IXXI} \\
$\Psi'\otimes$0111 & \texttt{X$\otimes$IXXX} \\
$\Psi'\otimes$1000 & \texttt{Z$\otimes$XIII} \\
$\Psi'\otimes$1001 & \texttt{I$\otimes$XIIX} \\
$\Psi'\otimes$1010 & \texttt{X$\otimes$XIXI} \\
$\Psi'\otimes$1011 & \texttt{X$\otimes$XIXX} \\
$\Psi'\otimes$1100 & \texttt{Z$\otimes$XXII} \\
$\Psi'\otimes$1101 & \texttt{X$\otimes$XXIX} \\
$\Psi'\otimes$1110 & \texttt{XZ$\otimes$XXXI} \\
$\Psi'\otimes$1111 & \texttt{Z$\otimes$XXXX}
\end{tabular}
\caption{Action required to correct the data qubit $\Psi'$ vs
measured value of ancilla qubits. Note that the X-operations
simply reset the ancilla.}
\label{table:one}
\end{table}

When simulating the QEC cycle, the circuit of
fig.~\ref{figure:5qecboth2}b was used to keep the analysis
architecture independent. Each gate was modelled as taking the
same time, allowing the time $T$ to be made an integer such that
each gate takes one time step. Gates were furthermore simulated as
though perfectly reliable and errors applied to each qubit
(including idle qubits) at the end of each time step. The
rationale for including idle qubits is that in an LNN architecture
physical manipulation of some description is required to decouple
neighboring qubits which inevitably leads to errors.

Two error models were used --- discrete and continuous. In the
discrete model a qubit can suffer either a bit-flip (X),
phase-flip (Z) or both simultaneously (XZ). Each type of error is
equally likely with total probability of error $p$ per qubit per
time step. The continuous error model involves applying
single-qubit unitary operations of the form
\begin{equation}
\label{continuous} U_{\sigma} =\left(
\begin{array}{cc}
\cos(\theta/2)e^{i(\alpha+\beta)/2} & \sin(\theta/2)e^{i(\alpha-\beta)/2} \\
-\sin(\theta/2)e^{i(-\alpha+\beta)/2} &
\cos(\theta/2)e^{i(-\alpha-\beta)/2}
\end{array} \right)
\end{equation}
where $\alpha$, $\beta$, and $\theta$ are normally distributed
about 0 with standard deviation $\sigma$.

Both the single qubit and single logical qubit (5 qubits) systems
were simulated. The initial state
\begin{equation}
\label{state}
|\Psi\rangle=\frac{5}{13}|0\rangle+\frac{12}{13}|1\rangle
\end{equation}
was used in both cases as $|\langle\Psi|X|\Psi\rangle|^{2}\cong
0.5$, $|\langle\Psi|Z|\Psi\rangle|^{2}\cong 0.5$, and
$|\langle\Psi|XZ|\Psi\rangle|^{2}= 0$ thus allowing each type of
error to be detected. Simpler states such as $|0\rangle$,
$|1\rangle$, $(|0\rangle+|1\rangle)/\sqrt{2}$, and
$(|0\rangle-|1\rangle)/\sqrt{2}$ do not have this property. For
example, the states $|0\rangle$ and $|1\rangle$ are insensitive to
phase errors, whereas the other two states are insensitive to bit
flip errors. Let $T_{wait}$ denote the duration of the wait stage.
Note that the total duration of the encode, decode, measure and
correct stages is 14. In the QEC case the total time
$T=T_{wait}+14$ of one QEC cycle was varied to determine the time
that minimizes the error per time step defined by
\begin{equation}
\label{step error} \epsilon_{step}=1-\sqrt[T]{1-\epsilon_{final}}
\end{equation}
where $\epsilon_{final}=1-|\langle\Psi'|\Psi\rangle|^{2}$ and
$|\Psi'\rangle$ is the final logical qubit state. An optimal time
$T_{opt}$ exists since the logical qubit is only protected during
the wait stage and the correction process can only cope with one
error. If the wait time is zero, extra complexity has been added
but no corrective ability. Similarly, if the wait time is very
large, it is almost certain that more than one error will occur,
resulting in the qubit being destroyed during the correction
process. Somewhere between these two extremes is a wait time that
minimizes $\epsilon_{step}$. Table~\ref{table:two} shows
$T_{opt}$, $\epsilon_{step}$ and the reduction in error
$\epsilon_{step}/p$ versus $p$ for discrete errors.
Table~\ref{table:three} shows the corresponding data for
continuous errors. Note that, in the continuous case, the single
qubit $p$ has been obtained via 1-qubit simulations and a 1-qubit
version of equation~\ref{step error}.
\begin{table}
\begin{tabular}{c|c|c|c}
$p$ & $T_{opt}$ & $\epsilon_{step}$ & $\epsilon_{step}/p$ \\
\cline{1-4}
$10^{-2}$ & 25 & $1.7\times 10^{-2}$ & $1.7\times 10^{0}$ \\
$1.6\times 10^{-3}$ & 40 & $1.6\times 10^{-3}$ & $1.0\times 10^{0}$ \\
$10^{-3}$ & 50 & $8.4\times 10^{-4}$ & $8.4\times 10^{-1}$ \\
$10^{-4}$ & 150 & $3.1\times 10^{-5}$ & $3.1\times 10^{-1}$ \\
$10^{-5}$ & 750 & $1.1\times 10^{-6}$ & $1.1\times 10^{-1}$ \\
$10^{-6}$ & 1500 & $3.2\times 10^{-8}$ & $3.2\times 10^{-2}$ \\
$10^{-7}$ & 6000 & $1.1\times 10^{-9}$ & $1.1\times 10^{-2}$ \\
$10^{-8}$ & 10000 & $2.0\times 10^{-11}$ & $2.0\times 10^{-3}$
\end{tabular}
\caption{Probability per time step $\epsilon_{step}$ of a discrete
error when using 5-qubit QEC vs physical probability $p$ per qubit
per time step of a discrete error.}
\label{table:two}
\end{table}

\begin{table}

\begin{tabular}{c|c|c|c|c}
$\sigma$ & $T_{opt}$ & $p$ & $\epsilon_{step}$ & $\epsilon_{step}/p$ \\
\cline{1-5}
$10^{-1}$ & $2.5\times 10^{1}$ & $5.9\times 10^{-2}$ & $6.9\times 10^{-3}$ & $1.2\times 10^{-1}$ \\
$10^{-2}$ & $2.5\times 10^{2}$ & $5.9\times 10^{-3}$ & $1.4\times 10^{-5}$ & $2.4\times 10^{-3}$ \\
$10^{-3}$ & $2.5\times 10^{3}$ & $6.0\times 10^{-4}$ & $1.3\times 10^{-8}$ & $2.2\times 10^{-5}$ \\
$10^{-4}$ & $2.5\times 10^{4}$ & $6.0\times 10^{-5}$ & $1.0\times 10^{-11}$ & $1.7\times 10^{-7}$ \\
$10^{-5}$ & $2.5\times 10^{5}$ & $6.0\times 10^{-6}$ & $7.2\times 10^{-15}$ & $1.2\times 10^{-9}$
\end{tabular}
\caption{Probability per time step $\epsilon_{step}$ of a discrete
error when using 5-qubit QEC vs standard deviation $\sigma$ of
continuous errors.}
\label{table:three}
\end{table}

An enormous range of threshold error rates $p$ exist in the
literature. These start at a very pessimistic $p=10^{-8}$
\cite{Knil96b} and go up to a very optimistic $p=2\times10^{-3}$
\cite{Zalk96}. The first thing that can be noted from the discrete
simulation data of Table~\ref{table:two} is that the LNN threshold
$p = 1.6 \times 10^{-3}$ is comparable to the most optimistic
previous estimate which was made using 7-qubit fault tolerant QEC
with errors applied only after gate operations and not to idle
qubits. The error rate $p = 1.6 \times 10^{-3}$ should not however
be thought of as the allowable operating error rate of a physical
quantum computer as precisely no improvement in error rate is
achieved when using QEC. If an error rate improvement of a factor
of 10 or 100 is desired when using QEC then $p = 10^{-5}$ or $p =
10^{-7}$ is required respectively. Further work is required to
determine the error rate improvement required to allow robust
implementation of large scale quantum algorithms with a reasonable
number of error correction qubits.

For continuous errors, there is no true threshold. Even for very
large random unitary rotations an improvement is still gained by
using the LNN QEC circuit. In this case, provided gates can be
implemented such that the angles associated with the continuous
error model are of order $10^{-2}$, an improvement in error rate
of at least a factor of 100 can be achieved.

Further work is required to determine whether the discrete or
continuous error model or some other model best describes errors
in physical quantum computers.

In conclusion, we have presented an efficient circuit for 5-qubit
QEC on an LNN architecture and simulated its effectiveness against
both discrete and continuous errors. It was found that, for the
discrete error model, if error correction is to provide an error
rate reduction of a factor of 10 or 100, the physical error rate
$p$ must be $10^{-5}$ or $10^{-7}$ respectively. For the
continuous error model, it was acceptable for error angles to have
a standard deviation of up to $10^{-2}$ radians as using QEC still
gives an error rate improvement better than a factor of 100.

Further simulation is required to determine the error thresholds
associated with 1- and 2-qubit LNN error-corrected gates.

\bibliography{references} 

\widetext

\end{document}